\documentclass{article}
\RequirePackage{amsthm,amsmath,amsfonts,amssymb,color}
\RequirePackage[authoryear]{natbib}

\usepackage[utf8]{inputenc}
\usepackage[english]{babel}
\usepackage{graphicx,url}
\usepackage[authoryear]{natbib}

\usepackage{csquotes}
\usepackage{booktabs} 

\newcommand{\hide}[1]{} 
\begin{document}


\title{Imputation Strategies for Rightcensored Wages in Longitudinal Datasets}
\author{Johannes Ludsteck\footnote{Corresponing author. IAB Nuremberg}, Joerg Drechsler\footnote{IAB Nuremberg, University of Mannheim, Ludwig-Maximilians-University Munich, University of Maryland. We thank Andreas Moczall and Alexandra Schmucker for helpful discussions. All remaining errors are our owns.}}
\date{}

\maketitle

\begin{abstract}
Censoring from above is a common problem with wage information as the reported wages are typically top-coded for confidentiality reasons. In administrative databases the information is often collected only up to a pre-specified threshold, for example, the contribution limit for the social security system. While directly accounting for the censoring is possible for some analyses, the most flexible solution is to impute the values above the censoring point. This strategy offers the advantage that future users of the data no longer need to implement possibly complicated censoring estimators.
However, standard cross-sectional imputation routines relying on the classical Tobit model to impute right-censored data have a high risk of introducing bias from uncongeniality \citep{Meng1994} as future analyses to be conducted on the imputed data are unknown to the imputer. Furthermore, as we show using a large-scale administrative database from the German Federal Employment agency, the classical Tobit model offers a poor fit to the data. 

In this paper, we present some strategies to address these problems. Specifically, we use leave-one-out means as suggested by \citet{CardEtAl2013} to avoid biases from uncongeniality and rely on quantile regression or left censoring to improve the model fit. We illustrate the benefits of these modeling adjustments using the German Structure of Earnings Survey, which is (almost) unaffected by censoring and can thus serve as a testbed to evaluate the imputation procedures.

    


  \vspace{1em}


\vspace{1em}


\end{abstract}


\section{Introduction}
Censoring from above is a notorious problem when analyzing wage data. In many administrative data sources the wage information is only collected up to an administrative limit such as the maximum taxable amount or the maximum contribution to the social security system. But even if information on the entire wage distribution is collected,  statistical agencies typically apply top-coding strategies when disseminating wage information to the public to protect the confidentiality of the respondents. With top-coding, values above a predefined threshold are not revealed. Values above the threshold are typically replaced with the value of the threshold or sometimes with the mean of all units affected by top-coding. 

Examples for both types of censoring are plentiful: In the Current Population Survey (CPS), 
the U.S. Census Bureau has recently adjusted its rules. It now top-codes the top three percent of reported earnings in each month \citep{CPS_SDC_update}. 
While three percent of the data might not seem much, the rate will be much higher when analyzing specific subgroups of the data, such as highly educated respondents. Other prominent surveys that use top-coding include the
American Community Survey (ACS), the Panel Study of
Income Dynamics (PSID), or the Survey of Income and Program Participation (SIPP). Outside the U.S. top-coding is used for example in the UK Household Logitudinal Study and the Quarterly
Labour Force Survey in the UK. 


A prominent example of censoring in administrative data motivated the research presented in this paper. The Employment History Data (BeH) at the Institute for Employment Research is a very rich administrative data source containing detailed employment information such as duration of employment, earnings, occupation, industry of the employer etc. for all German employees covered by the Social Security System. It has been the basis of various influencial papers (see, for example, \citet{CardEtAl2013,Wachter2006,Schmieder2016,Schmieder2012,DustmannEtAl2009,Schonberg2014,DustmannEtAl2016,DustmannSchoenberg2012}.)\hide{\color{red} Hans, ich hoffe hier gibt es ein bisschen mehr, könntest du hier ergänzen?} The data are based on notifications that every employee in Germany has to provide regarding his employees on a regular basis. Since the administrative purpose of this database is to set the social security payments, all wage information is only collected up to the contribution limit. This implies that the wage information is censored from above. Censoring shares range between about 10 and 12 percent (depending on the year) averaging across all full-time employed men but exceed 30 percent for men with college degree.

Similar problems arise in the Austrian Social Security Database, which also only provides wage information up to the contribution limit. In the U.S., the Earnings Public-Use Microdata File published by the Social Security Administration based on social security tax records are censored at the maximum taxable earnings level of the social security \cite{SSA_Cens}. 

\hide{The problem of wage censoring is especially severe for analyses of wage inequality since movements at the upper end of the income distribution affect mainly the high qualified became more important in the recent past. {\color{red} könntest du hier eine Zitation ergänzen und vielleicht generell noch etwas erläutern, warum das mit der Zensierung so ein Problem ist?}}

Two general strategies are commonly applied to deal with the censoring problem. Either the censoring is directly taken into account when analyzing the data or a two-stage procedure is employed in which all censored values are imputed first before applying standard analysis procedures using the imputed data (potentially accounting for the extra uncertainty from imputation). For the first approach, different strategies are applied depending on the type of analysis to be conducted. If wages are treated as the dependent variable in a regression context, the most common approach is to replace the linear regression by Tobit models \citep{tobin1958estimation}. If the censored variable is used as a predictor, most researchers follow \citet{Chow1979} and \citet{AndersonEtAl1983} who propose to interact the censored regressor with a dummy which indicates whether the observation is censored and to add this dummy and the interaction term to the model. As discussed for example in \cite{Jones1996}, the second approach should generally be avoided as it can introduce bias in the estimated regression coefficient of the censored variable, which may translate into biases in the coefficients of the other variables (depending on their partial correlations with the censored regressor).

But even the Tobit model, which provides unbiased results as long as the model assumptions are fulfilled, has the drawback that it can only be used to obtain estimates for the linear regression model. If interest lies on other aspects such as studying income inequality, the Tobit model will not be helpful. Besides, the Tobit model has the disadvantage that compared to standard OLS the set of regression diagnostics is very limited and model selection procedures can be time consuming for large datasets since the parameters cannot be obtained in closed form.  

\hide{two major drawbacks: 
\begin{enumerate}
    \item Model selection procedures are slow for large data sets since regression parameters can no longer be obtained in closed form and the required numeric
    optimization routines tend to be time consuming. They become slow and cumbersome especially for models with large sets of Dummy variables (e.g. for industries, occupations and regions).
    \item The rich set of Least Squares diagnostic tools (outlier statistics, $R^2$, resdidual-based tests, Ramsey RESET test, added-variable plots (based on the Frisch-Waugh theorem) etc.) is not directly available or applicable.
\end{enumerate}
}

Thus, the imputation approach is often preferred in practice as it offers full flexibility regarding the type of analysis conducted on the imputed data. This strategy has been used in various contexts, most importantly when studying wage inequality, e.g in \cite{DustmannEtAl2009,CardEtAl2013} (we provide a detailed literature review in Section \ref{sec:imp_review}). An additional important advantage of the imputation approach is the possibility of reusing the imputed variable for other research projects. This can reduce the burden for future projects that otherwise always have to come up with their own strategy how to deal with the censoring problem. Furthermore, given the larger potential benefits it justifies some extra efforts to carefully design and evaluate the model used for imputation. 

However, there is an important caveat to this approach. The imputed values will only reflect those relationships that were built into the imputation model. \citet{Meng1994} coined the term \textit{uncongeniality} to describe the situation if the modeling assumptions differ between the imputer and the analyst. If the imputer and the analyst are different individuals, uncongeniality is almost inevitable. Uncongeniality is especially problematic if variables or interaction terms that are included in the analysis model are not included in the imputation model. The regression coefficients of these variables will be attenuated after imputation unless the implicit assumption of the imputation model is satisfied that these variables are no longer correlated with the dependent variable given the variables included in the model.

A general recommendation in the imputation literature is therefore to always use inclusive models based on a rich set of predictors. The more of the variability of the dependent variable can be explained by the predictors the smaller the possible attenuation bias for any variables not included in the model. In the context of wage regressions this implies that person level and establishment level fixed effects should always be included in the imputation model. The inclusion of these effects is important for two reasons: First, the effects control for all time-invariant individual and establishment level effects avoiding omitted variable bias and substantially improving the model fit. For example \cite{AbowdEtAl1999} find that the $R^2$ in a wage regression improves from about 0.4 to 0.9 if these fixed effects are included. Second, including the establishment level effects will implicitly control for all regional and industry level effects as well. Since establishments rarely move their geographical location or change their main activity to the extend that they would be classified as belonging to a different industry, the regional and industry effects are simple aggregates of the establishment effects and are thus already taken into account by the inclusion of the establishment level effects.     

However, directly including the fixed effects as dummies is often not feasible in practice because of the large number of parameters that would need to be estimated. Furthermore, the commonly applied within-transformation strategy, which helps to reduce the number of parameters, can also not be employed, as the necessary average wages on the individual and establishment level are not available due to censoring. A convenient strategy to address this problem was first discussed in \cite{CardEtAl2013}. They suggest to approximate the fixed effects by leave-one-out-means (LOOMs). 

\begin{figure}
    \caption{Density of the log wages after imputation based on the Tobit model proposed in \citet{CardEtAl2013}. The figure shows an unrealistic drop in the wage distribution directly at the censoring limit (indicated by the red vertical line).}\label{fig:tob-imp}
    {\begin{center}
    \includegraphics[width=0.8\textwidth]{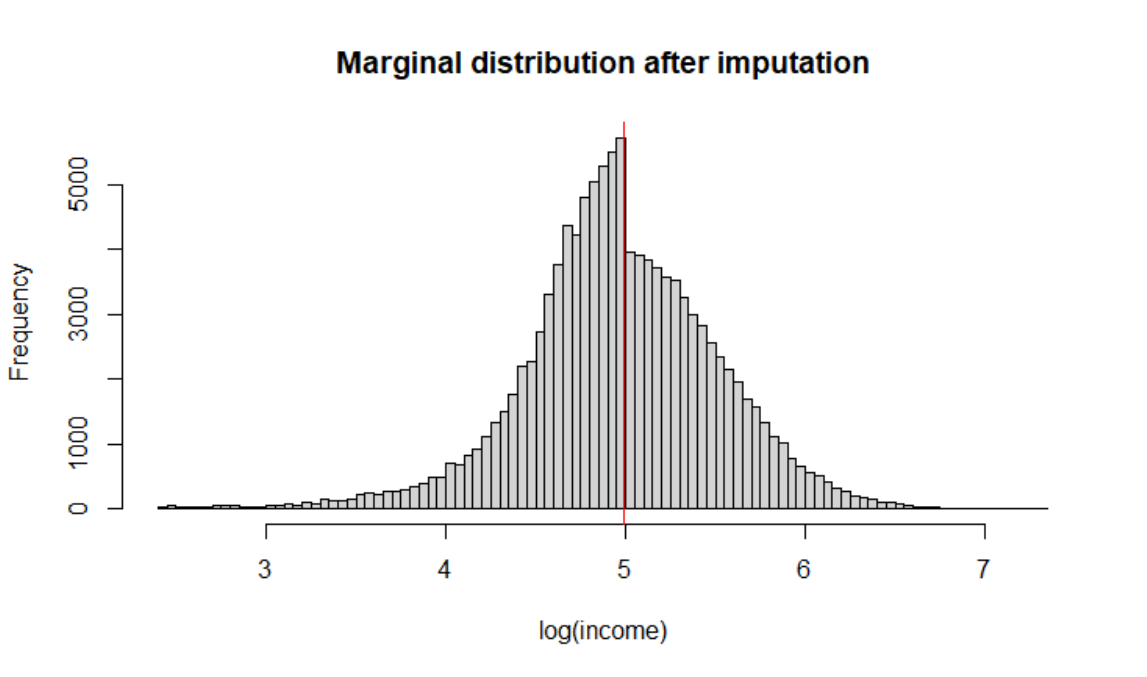}
    \end{center}}
    Data Source: BeH, Year 2010
    
    Subsample: Men aged 30-45 years, Western Germany, high qualification (college or technical college)
\end{figure}
However, as illustrated in Figure \ref{fig:tob-imp}, even with this strategy imputations based on the Tobit model result in unrealistic patterns around the censoring limit. We demonstrate in this paper that the problem arises since the assumption of constant regression coefficients is violated. We propose three different strategies to address this problem: (1) fitting a doubly censored Tobit model which reduces the bias by limiting the contribution of observations in the lower tail of the distribution of the dependent variable, (2) fitting a truncated quantile regression model and using estimated coefficients close to the censoring limit, (3) fitting a truncated quantile regression model and extrapolating the estimated regression coefficients using weighted ridge regression.

In addition to the BeH, we also use the Structure of Earnings Survey (Verdienststrukturerhebung, VSE) of the German Federal Statistical Office \citep{hafner2008german} to assess the feasibility of the different imputation approaches. The VSE offers an ideal test bed for evaluation since income information in the VSE is censored for less than 1\% of the data.

The remainder of the paper is organized as follows: In Section \ref{sec:imp_review} we review the previous literature on imputation strategies for right-censored wages. Section \ref{sec:data} introduces the two datasets that we use in our evaluation, the BeH and the VSE. In Section \ref{sec:Tobit_problems} we illustrate the violation of the constant coefficient assumption before we discuss the three proposed adjustment strategies in more detail in Section \ref{sec:imp_models}.  In Section \ref{sec:evaluation} we present the results of our evaluation. The paper concludes with some final remarks. 

\hide{However, they do not account for the fact that these LOOMs will also be affected by censoring. \hide{As we show in Section XX, not accounting for the censoring will still introduce bias in the imputed data (I HOPE WE CAN SHOW THIS).} To overcome this problem, we propose a two-stage imputation routine, in which censored values are replaced with imputed values based on a model without fixed effects on the first-stage. These imputed values are used as input for the second stage, which includes the LOOMs computed based on the imputed values from the previous round. We also identify another problem: simply using all information below the contribution limit, will introduce bias in the imputed values generating a heap in the imputed data immediately above the contribution limit. Since this heap occurs in two different datasets that we use in our evaluations, we believe that this is a general problem, which is not limited to our data. We demonstrate that the problem arises since the assumption of constant regression coefficients is violated and propose two alternative methods to cope with it: (1) A doubly censored Tobit model which reduces the bias by reducing the contribution of observations in the lower tail of the distribution of the dependent, and (2) estimating the regression coefficients near the censoring limit using censored quantile regressions.}

\section{Review of previous imputation strategies for dealing with right-censored wage data}\label{sec:imp_review}
Simplistic imputation strategies such as setting all values above the censoring threshold to the threshold times some constant factor or to the expected mean above the threshold have been used in many papers (see, for example, \citet{katz1992changes,lemieux2006increasing,autor2008trends}). Such strategies will obviously be problematic if an imputed version of the wage variable should be released as they would lead to biased results for almost all analyses conducted using the imputed variable. Beyond those simplistic imputations, some researchers only used the observed marginal income distribution to impute censored values. They typically either assume that income follows a Pareto distribution \citep[among others]{fichtenbaum1988truncation,bishop1994truncation, bonke2015, armour2016} or model income based on the beta distribution of the second kind \citep{jenkins2011, schluter2023spatial}. In fact, \citet{schluter2023spatial} find that their imputation approach outperforms the classical imputation strategy based on Tobit models discussed below. However, a major drawback of all imputation approaches discussed so far is that they do not model the relationship to the other variables in the data. This is unproblematic as long as interest only lies in the marginal distribution, for example, when studying wage inequality. If the goal is to fit regression models using the imputed variable, results will be biased as all coefficients will be attenuated towards zero due to the implicit independence assumption for the imputed values. 

To the best of our knowledge all previous regression based imputation approaches relied on the Tobit model. In 1987, \citeauthor{ham1987unemployment} used the Tobit model to impute wages above the censoring limit in the Canadian Employment and Immigration Longitudinal Labour Force File. However, they used the predicted values from the model for their  imputations, which underestimates the variability in the wage distribution leading to biases in the obtained regression results. Stochastic regression imputations were used for example in \citet{gartner2005imputation,haider2006life,DustmannEtAl2009,LehmerLudsteck2011,Ludsteck2014,LehmerLudsteck2014,gobel2013personnel}. Some authors also used multiple imputation to fully account for the uncertainty in the imputed values \citep{gartner2005analyzing,jensen2010estimating} and introduced strategies to deal with heteroscedasticity in the conditional wage distribution \citep{buttner2008multiple,brucker2014migration}. 

In most of these papers, the imputation step was only a preliminary data preperation step before conducting the actual analysis of interest, i.e., in these papers the imputer and the analyst were typically the same person. Thus, the authors could ensure that their imputation model was congenial to their analysis of interest. However, as discussed earlier, the goal of the project that motivated this research was to be able to offer an imputed wage variable that researchers can use
irrespective of their analysis goals. To approximately achieve this, it is important to find a model that explains most of the variability of the dependent variable. The logic is simple: if the variables included in the model already explain almost all of the variability of the dependent variable, it becomes irrelevant that some variables that are used in the analysis later have not been included. Given the variables included in the imputation model they cannot have much explanatory power for the dependent variable, i.e., their conditional correlation has to be close to zero. As discussed earlier, the explanatory power of a wage model can be improved substantially by including person level and establishment level fixed effects. However, this is typically infeasible in practice due to the large number to predictors that would need to be estimated. \citet{CardEtAl2013} suggested a convenient solution to this problem by approximating the fixed effects by including leave-one-out means (LOOMs). We will review the methodological details in the next section.

While the strategy of \citet{CardEtAl2013} helps to limit the uncongeniality problem, Figure \ref{fig:tob-imp} illustrates that it still does not guarantee reliable imputations above the censoring limit. After introducing the two data sources that we will use in the remainder of the paper, we provide reasons for the poor model fit and propose several strategies to mitigate the problem in Section \ref{sec:Tobit_problems}.

\section{Data sources}\label{sec:data}
\subsection{The German Employee History (BeH)}
The Employee History (Betriebshistorik, BeH) of the IAB is an administrative database that covers event history data (at daily precision) on \textit{all} employees liable to social security\footnote{This amounts to about 80 percent of the
German workforce.} since 1975. It contains information on wages and other important individual
characteristics (gender, age, schooling and occupational qualification level, job status) as well as occupation, industry and establishment identifiers. Its wage information is highly reliable since the data are collected in order to determine unemployment, health and pension insurance contributions and misreporting is punished by severe financial penalties. Wages are censored, however, at the contribution assessment ceiling, which amounts to about 10 -- 13 percent of all full-time employed and exceeds 30 percent for the high-qualified (those with a Bachelor or Master degree).
See \cite{SchmuckerEtAl2023} for a detailed description of the dataset.

\subsection{The VSE of the German Federal Statistics Office} 
The Structure of Earnings Survey (Verdiensstrukturerhebung, VSE) is a large survey conducted by the German Federal Statistical Office comprising earnings, working time and job status information on more than one million employees from more than 50.000 companies. Its representativness is ensured by design-based survey sampling and compulsory participation. The VSE is well suited for the  evaluation of our imputation models since wage censoring is limited to the top 1\% of the wage distribution
and the definitions of important personal characteristics and wages exactly match their counterparts in the BeH. A one-to-one evaluation of our models is, however, impossible due to some deficiencies of the VSE. First, it is a cross-sectional survey (conducted every four years since 2006). Thus, it is not possible to include LOOMs in the imputation model as proposed by \citet{CardEtAl2013}. Furthermore, important characteristics based on biographical information such as the share of non-employment or minor employment episodes in the gross employment history are not available. Finally, the VSE 2010 (this is the wave used for our evaluation exercises) does not cover the entire population of establishments in Germany. It excludes establishments with less than 10 employees and three small sectors (sections A, T and O, agriculture, private households and extraterritorial organizations.\footnote{The three-digit codes of these sectors are WZ08$\le$50 and WZ08$\ge$970 and 841 $\ge$ WZ08 $\le$ 859.})  We mimic this by applying the same restrictions\footnote{Furthe restrictions are: (1) spells which do not intersect the entire October 2010 from the BeH since the VSE relates to emploment and wages in October 2010 and (2) civil servants are dropped from the VSE since the are not covered by the BeH.} to the BeH subsample that is used in our evaluation study described in Section \ref{sec:evaluation}. 

\begin{figure}
    \caption{Comparison of the Densities of Log Wages}
    {\centering
    \includegraphics[width=0.51\textwidth]{%
    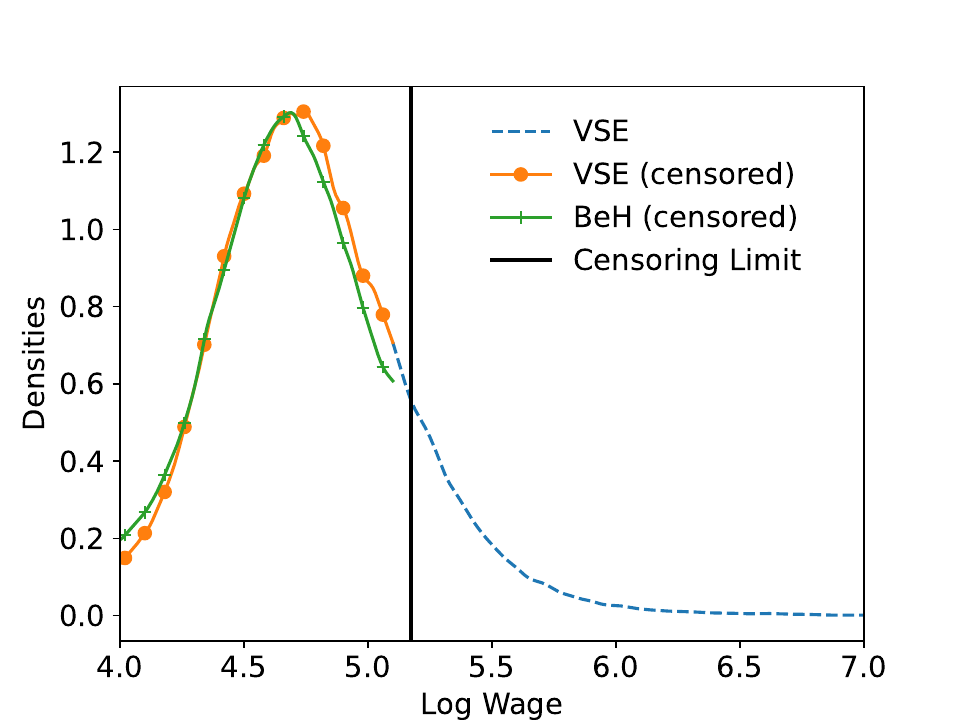}
    \includegraphics[width=0.51\textwidth]{%
    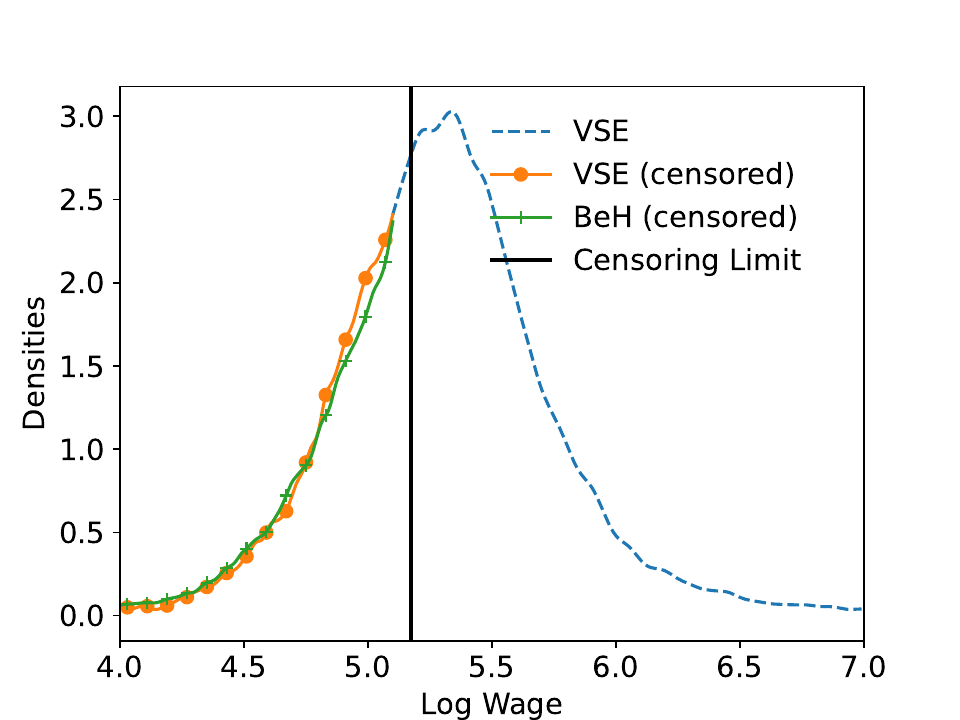}
    }

    Legend: Left hand side: Men aged 30-64 years, Western Germany, medium qualification (completed apprenticeship).
    
    Right hand side: Men aged 30-64 years, Western Germany, high qualification (college or technical college).
    \label{fig:comp-densities-beh-vse}
\end{figure}

To check the similarity of the datasets we compared the frequency distribution of the age variable and computed kernel density estimates of log daily wages for several subsamples. The densities are sufficiently similar in some of the subsamples (men aged 30--64 years, working in Western Germany, with medium and high qualification), but show noteworthy deviations for others (mainly women working in Eastern Germany). Therefore most evaluations are restricted to men working in Western Germany.\footnote{Low-qualified employees are excluded from our analysis due to the  small censoring shares  (below 5 percent or smaller).}
Table \ref{tab:comp-beh-vse-age} compares the distribution of the age groups for the entire datasets while Figure \ref{fig:comp-densities-beh-vse} depicts kernel densities of log wages for those subsamples for which the distributions of uncensored wages are sufficiently similar to allow reliable comparisons.

\begin{table}[t]
    \caption{Frequencies of Age Groups in the VSE and BeH}
    {\centering
    \begin{tabular}{llrrr}
\toprule
{} & Age Group &  Percent VSE &  Percent BeH &  Difference \\
\midrule
0 &  [14, 25) &         6.56 &         6.58 &       -0.03 \\
1 &  [25, 35) &        22.93 &        21.98 &        0.96 \\
2 &  [35, 45) &        27.77 &        30.09 &       -2.32 \\
3 &  [45, 55) &        30.47 &        30.78 &       -0.31 \\
4 &  [55, 65) &        12.28 &        10.57 &        1.71 \\
\bottomrule
\end{tabular}

    \label{tab:comp-beh-vse-age}
    }
    
    Samples: All dependent (female and male) workers
\end{table}

Inspection of the frequency table and the kernel density plots suggests that the  similarity of the data sets is sufficient for our purpose.

\section{Limitations of the Tobit imputation model and possible strategies to tackle them}\label{sec:Tobit_problems}
\begin{figure}
    \caption{Profiles of the Coefficients from Censored Quantile Regressions}
    \label{fig:coeff-stability-vse}
      \includegraphics[width=1.1\textwidth]{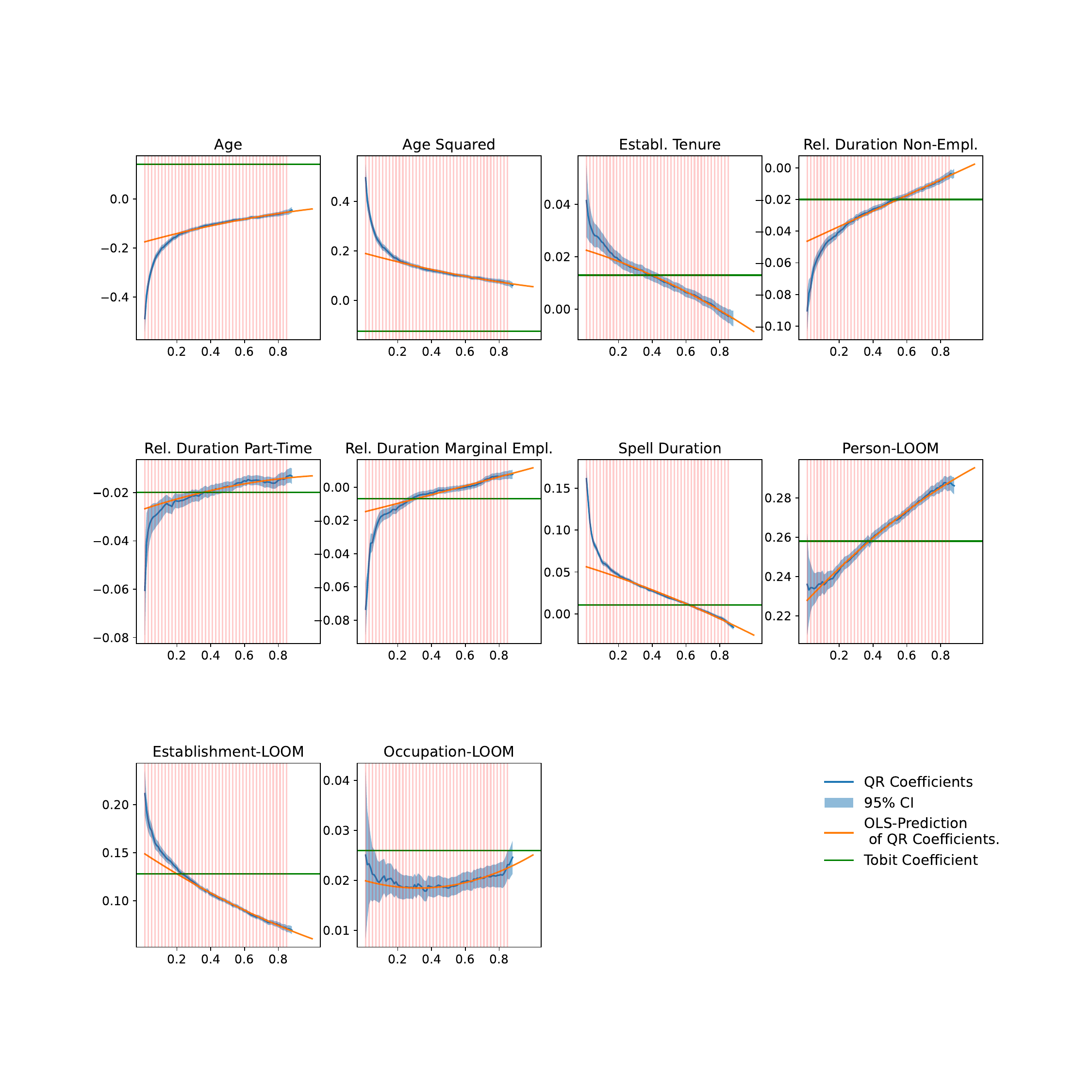}
    {\footnotesize Source: Own computations

  The light-red dashing indicates the uncensored range.

  Data Source BeH, subsample: Medium-qualified men (completed apprenticeship),
    working in western Germany, age group 30 to 44 years.}
\end{figure}

A strong limitation of the Tobit model is its inflexible parametric form. Most notably, the model assumes constant regression coefficients for the entire conditional wage distribution. To assess the validity of this assumption, we fit quantile regression models to the BeH. Since we stratify the data by various variables and apply separate imputation models for each stratum in the final application, we use the same imputation cells for the assessment. Figure \ref{fig:coeff-stability-vse} contains exemplary results from one of these cells (males between 30 and 45 with medium qualification level working in Western Germany). The results for the other imputation cells showed similar patterns. The figure plots the estimated regression coefficients as a function of the conditional wage quantile for selected regressors. The horizontal solid line in each plot shows the coefficient obtained using Tobit regression while the vertical lines indicate the support of the data (the yellow dashed line will be explained later). The figure demonstrates a clear violation of the constant coefficient assumption. It also shows that the coefficients obtained from the Tobit regression are often very far from the actual coefficients in the area of the wage distribution for which imputations are desired. 

Given the obvious violation of the constant parameter assumption, we propose three strategies to address this problem:
\begin{enumerate}
\item Fitting a doubly censored Tobit model which reduces the bias by limiting the contribution of observations in the lower tail of the distribution of the dependent variable.
\item Fitting a truncated quantile regression model and using estimated coefficients close to the censoring limit
\item Fitting a truncated quantile regression model and extrapolating the estimated regression coefficients using weighted ridge regression.
\end{enumerate}

The first approach is attractive as it offers a simple adjustment to the standard Tobit model. Since most software packages for censored regression allow for doubly censored variables, the approach is very simple to implement. It is motivated by the fact that the estimated coefficients in Figure \ref{fig:coeff-stability-vse} change most dramatically below the 20th quantile and thus censoring from below should stabilize the estimated coefficients. The second approach improves over the standard Tobit model as the estimated coefficients at the censoring limit will likely be closer to the true coefficients beyond the censoring limit. However, the approach relies on the assumption that the coefficients are stable beyond the censoring limit. The final approach uses a model to predict the regression coefficients beyond the censoring limit and uses these predictions for the imputation. The yellow dashed line in Figure \ref{fig:coeff-stability-vse}, which shows the predicted coefficients from the model illustrate that this approach seems promising. The regression line closely follows the estimated coefficients from the quantile regression model. However, as we will see in our evaluations, the approach strongly relies on the validity of the modeling assumptions and can introduce bias if these assumptions are not met.

\section{The imputation models}\label{sec:imp_models}
In this section, we review the details of the different imputation models that we consider for imputing wages beyond the censoring threshold. We start with the classical Tobit model including LOOMs as suggested in \citet{CardEtAl2013} before presenting the various adjustments that we propose to deal with the poor fit of the basic model.

\subsection{The Tobit model as implemented in \citet{CardEtAl2013}}

Similar to early works (e.g. \cite{DustmannEtAl2009,CardEtAl2013}) our baseline imputation model is based on the Tobit model. To maximize the explanatory power and to fully account for the hierarchical structure of the data (employment spells nested within individuals and individuals nested within establishments and within occupations), the ideal model would be given as
\begin{equation}
    w_{siet} = \min\big\{C_t, x_{siet} \, b_t + \mu_i + \eta_{et} + \omega_{ot} + u_{siet}\big\},
\end{equation}
where
\begin{itemize}
 \item $s,i,e,o,t$ denote identifiers for spells, persons, establishments, occupations and time, respectively. 
 \item  $w_{siet}$ denotes the natural logarithm of daily (pre-tax) wages, censored at the social contribution limit $C_t$.
  \item $C_t$ denotes the social contribution limit.\footnote{The censoring threshold is somewhat smaller (roughly 20 to 30 Euro) in Eastern Germany. We do not use an additional
  subscript ($C_{t,r}$) for sake of notational simplicity.}
  \item $x_{siet}$ denotes a row vector of predictors that vary over spells, persons, establishments and time.
  \item $\mu_i$ denote fixed person effects.
  \item $\tilde\eta_{et}$ denote time-varying fixed establishment effects.
  \item $\tilde\omega_{ot}$ denote time-varying fixed occupation effects.
  \item $u_{siet}$ denote residuals, which under the classical Tobit model assumption are normally distributed with zero mean and constant variance $\sigma^2$.
\end{itemize}

However, as discussed in the introduction, estimating the true fixed effects  $\mu_i$, $\eta_{et}$ and $\omega_{ot}$ by adding dummies for the respective groups is computationally infeasible due to the large number of groups (several thousand individuals contained in each estimation cell imply that thousands of coefficients would need to be estimated). Fitting this model would also  
yield biased estimates\footnote{Consistent estimation of the person-level dummies is infeasible since the coefficients are based on a small number of observations (on average less than 30 spells per person). Due to the nonlinearity of the Tobit estimator this inconsistency translates to all other coefficients, see, for example, \cite{Hsiao2003}, p. 194 and 243. A consistent method-of-moments estimator for Tobit models with large numbers of fixed effects was proposed by \cite{Honore1992}. This estimator is not suitable for imputation purposes since it irretrievably removes the fixed effects.}. 

We avoid this problem by following \cite{CardEtAl2013} who approximate the fixed effects by adding
the (spell-duration weighted) leave-one-out means (LOOMs) of daily wages as regressors. Formally, $\tilde{\mu}_{i, -s}$ is the duration-weighted mean over all spells of person $i$ except for spell $s$. Correspondingly $\tilde{\eta}_{et,-i}$ relates to all co-workers of person $i$ in establishment $e$ and year $t$. Finally, $\tilde{\omega}_{ot,-i}$ is the average wage of all workers in occupation $o$ (except $i$). A formal definition of the LOOMS is provided in Appendix \ref{sec:appendix-formulas}.

After substituting the fixed effects by their approximations the model can be written as
\begin{equation} \label{eq:est-imput}
w_{siet} = \min\big\{C_t, x_{siet} \, b_t
       + \tilde{\mu}_{i,-s} \, h_\mu + \tilde{\eta}_{et,-i} \, h_\eta + \tilde{\omega}_{ot,-i} \, h_\omega + u_{siet}\big\}
       .
       \end{equation}

To improve the fit, the model is estimated separately for subgroups of the register data which are obtained by
partitioning the dataset by year, gender, four age groups, four education groups and Eastern/Western Germany. 

The imputed wages $w_{siet}^I$ are computed as
\begin{equation} \label{eq:pred-imput}
 w_{siet}^I := x_{siet} \, \hat{b}_t + \tilde{\mu}_i \, \hat{h}_\mu + \tilde{\eta}_{et} \, \hat{h}_\eta + \tilde{\omega}_{ot} \, \hat{h}_\omega + \tilde{u}_{siet}.
 \end{equation}
for censored observations. The error term $\tilde{u}_{siet}$ is sampled from a truncated normal distribution with variance $\hat{\sigma}_{siet}^2$. The estimated variance $\hat{\sigma}_{siet}^2$ includes both the variance of residuals and of the coefficient estimates
\begin{equation}
\hat{\sigma}_{siet}^2  =  \hat{V}(x_{siet} \, \hat{b}_t + u_{siet}) 
                    = x_{siet} \, \hat{V}(\hat{b}_t) \, x_{siet}^\top+\hat{V}(u_{siet})
\end{equation}
Sampling the error term from a left-truncated distribution ensures that $w_{siet}^I > C_t$ after adding
 $\tilde{u}_{siet}$. Substitution of definition (\ref{eq:est-imput}) into the condition
$ w_{siet}^I > C_t$ and solving for $\tilde{u}_{siet}$ yields
\begin{equation}
\tilde{u}_{siet}  > C_t - x_{siet} \, \hat{b}_t - \tilde{\mu}_i \, \hat{h}_\mu -\tilde{\eta}_{et} \, \hat{h}_\eta - \tilde{\omega}_{ot} \, \hat{h}_\omega
\end{equation}
as the left-truncation threshold.

Note that computing the leave-one-out means directly from the data (as implemented in \citet{CardEtAl2013}) would imply that the means would be based on censored wages. This may generate considerable bias, especially for individuals with large shares of censored wages. To see this, consider the extreme case where all observations of an individual are censored. Under this scenario the leave-one-out mean  is the mean of the censoring thresholds $C_t$ for those years in which the individual was employed.
Since the  combination of the leave-one-out means  
explains a large share of the variance of the dependent variable, the predicted values (before adding $\tilde{u}_{siet}$) would be close to the average censoring limit. We mitigate this effect by obtaining an initial estimate for the censored wage by dropping $\tilde{\mu}_{i,-s}$, $\tilde{\eta}_{et,-i}$ and $\tilde{\omega}_{ot,-i}$ from equation (\ref{eq:est-imput}). The predicted values from this step are used to compute $\tilde{\mu}_{i,-s}$, $\tilde{\eta}_{et,-i}$ and $\tilde{\omega}_{ot,-i}$ for the final imputation. 

\subsection{A doubly censored imputation model}

As discussed earlier, implementing the imputation approach as described in the previous section introduces  artifacts (kinks and bumps) in the distribution of the imputed wages above the censoring threshold 
due to the variability of the regression coefficients with respect to the quantiles of the dependent variable.
Since the variability in the coefficients is highest below the 20th quantile of the wage distribution, a simple solution to the problem is to introduce artificial left censoring when estimating the model parameters for the imputation model. To achieve this, model (\ref{eq:est-imput}) is extended to
\begin{equation} \label{eq:est-imput2}
w_{siet} = \max\Big\{c_t, \min\big\{C_t, x_{siet} \, b_t
       + \tilde{\mu}_{i,-s} \, h_\mu + \tilde{\eta}_{et,-i} \, h_\eta + \tilde{\omega}_{ot,-i} \, h_\omega + u_{siet}\big\}
       \Big\},
       \end{equation}

where $c_t$ denotes the artificial lower censoring limit introduced to ensure stable estimates of the regression coefficients. We also experimented with other censoring limits than the 20th quantile. However, using this limit tended to give the best results.
  
\subsection{An imputation model based on quantile regression}
A natural extension to the Tobit model is the censored quantile regression (CQR) model which adapts the standard quantile regression model as proposed by \cite{KoenkerBasset1978} to models with censored dependent variables. \cite{Powell1986} showed that censoring can be tackled by exploiting the invariance of regression quantiles with respect to nonlinear monotonic transformations. This triggered a series of proposals 
\citep{Buchinsky1994,KhanPowell2001,BuchinskyHahn1998, ChernouzhukovHong2002} which aimed at improving the implementation of Powell's estimator.\footnote{See \cite{Fitzenberger1997} for an overview and further details.} 

We use the three-step estimator proposed by \cite{ChernouzhukovHong2002} because of its computational and statistical efficiency. It is based on a standard probit model explaining the censoring indicator by the model's exogenous variables, followed by two quantile regressions. The first two regressions are used to select the subsample of observations with uncensored prediction for the third regression which yields asymptotically efficient estimates.

An obvious disadvantage of CQR is that the coefficients can only be estimated up to the censoring threshold. This is especially problematic in our application as we are only interested in modeling wages above the threshold. We evaluated two strategies to address the problem. The first approach uses the estimated coefficients close to the threshold while the second approach -- described in more detail in the next section -- uses a model to extrapolate the coefficients beyond the censoring limit. With the first approach, wages are imputed using the following model:
\[  w_i^I= x_i \, b_{CQR}(q_C) + \epsilon,\]
where $b_{CQR}(q_C)$ denotes the regression coefficients from the CQR at the largest uncensored quantile ($q_C$) of the wage distribution and $\epsilon$ is 
a draw from a left-truncated normal distribution where the truncation
limit is $
\epsilon > C - x_i \, b(q_C)$. This approach effectively boils down to assuming constant coefficients on the right of the censoring limit and extrapolating the value obtained slightly below the limit.

\subsection{Extension of the quantile regression imputation: extrapolated regression coefficients}\label{sec:QCR_extrapolation}
The approach described in the previous section relies on the assumption that the regression coefficients stabilize at the censoring limit. To avoid this assumption an alternative strategy is to use the observed distribution of the regression coefficients to model the distribution beyond the censoring limit.

Specifically, we regress the estimated coefficient profiles on a quadratic ploynomial of the quantile (plus intercept) and use the predictions from this model to extrapolate into the censored range.
We avoid overfitting of the lower quantiles (below the tenth quantile) by employing regularized least squares regressions (with L2-penalty terms and penalty weight 0.002). 
Furthermore, we  weight the observations with their quantile distance from the censoring limit to approximate a local linear regression, i.e.,
$w_i = q_c - q_i$ where $q_c$ is the quantile of the censoring limit and $q_i$ is the quantile of the 
coefficient estimate.

The extrapolation yields quantile coefficient estimates $\tilde{b}(q_i)$  for a narrow grid of quantiles $q_i \in \{0.01, 0.02, \ldots, 0.99, 1.0\}$. For each observation $i$ with censored wage information let $q_i^{min}$ denote the smallest quantile such that $ x_i \, \tilde{b}(q_i) \ge C_t$, where $C_t$ is the year-specific censoring limit. Imputed wages are obtained by randomly drawing $u_i$ from the truncated uniform distribution in the range $\{q_i^{min}, q_i^{min}+0.01, \ldots, 1]$ and generating imputed values as
    \[ w_i^I= x_i \, \tilde{b}(u_i).\]


\section{Evaluation study}\label{sec:evaluation}
To evaluate the feasibilty of the different imputation approaches we heavily rely on the VSE. As mentioned earlier, the VSE has the major advantage that censoring is limited to the top 1\% of the wage distribution and thus the data can be used as a testbed to evaluate the different imputation strategies. The downside of the VSE is that it is a cross-sectional survey. This implies that the LOOMs cannot be included in the imputation models. Thus, for most of our evaluations we have to assume that all imputation models are affected similarly by the exclusion of the LOOMs (we avoid this problem in Section \ref{sec:BEH_evaluations}, where we directly compare the harmonized marginal imputed wage distribution of the BeH with the the wage distribution of the VSE).

We start this section by evaluating the feasibility of extrapolating the coefficients of the CQR to the censored region. In Section \ref{sec:eval_imp}, we directly evaluate the different imputation approaches using the VSE. Finally, we assess whether the proposed imputation strategy (including the selection strategy discussed in Section \ref{sec:eval_imp}) yields improved imputations for the BeH.

\subsection{Evaluation of the extrapolation of the quantile regression coefficients}

\begin{figure}
    \caption{Comparison of extrapolated quantiles and estimates obtained from the unrestricted data}
    {\centering
    \includegraphics[width=1.1\textwidth]{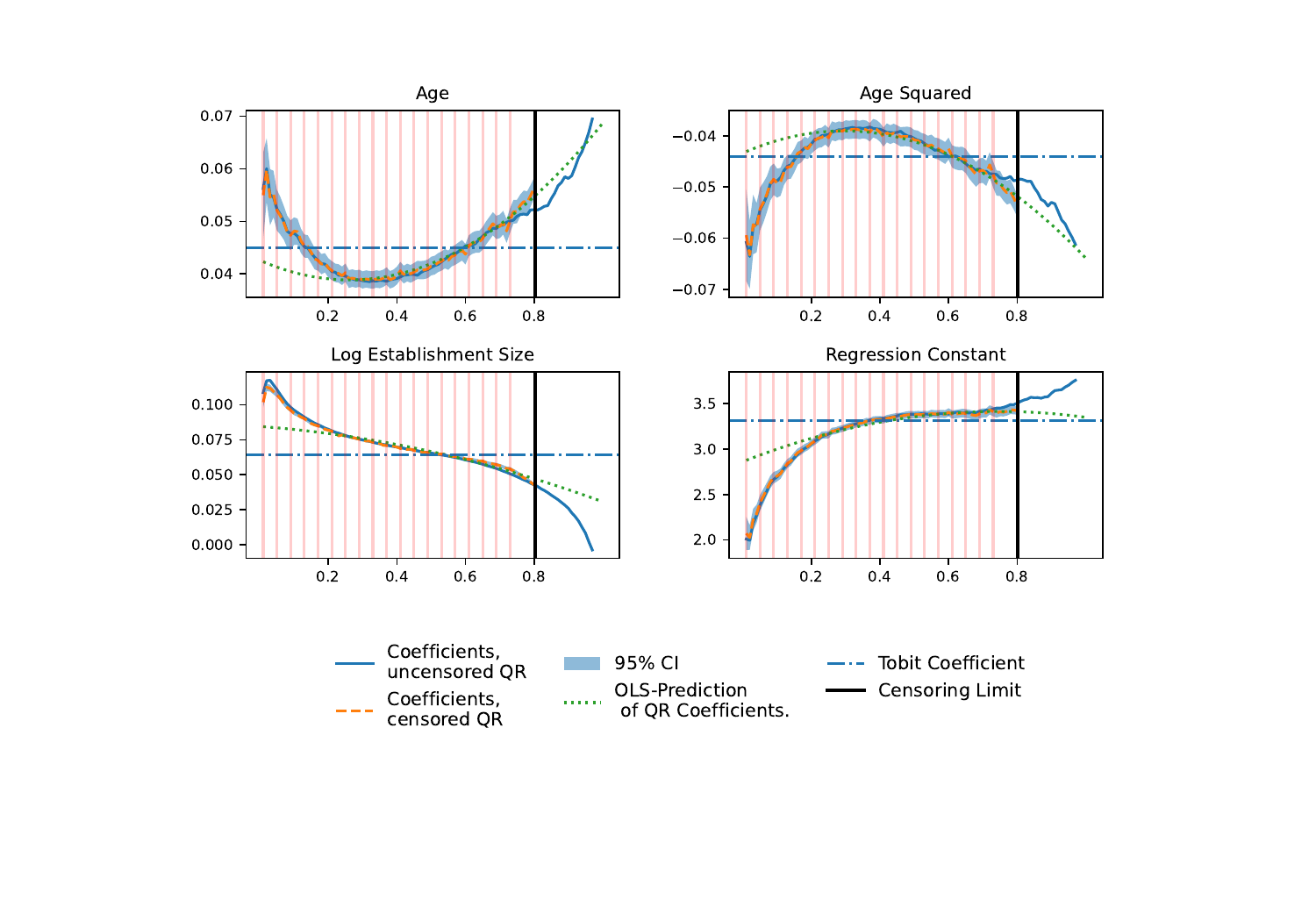}
    }
    \label{fig:coeff-stability-med-qual-vse}
\end{figure}

The extrapolation approach can be problematic if the fit of the quantile coefficient process by a least squares regression is poor, the censored range is large, or if the quantile coefficient path continues not smoothly or contains turning points in the censored range. We evaluated this approach by artificially censoring the VSE data using the censoring points in the BeH. We then fitted a CQR to these data and extrapolated the coefficients beyond the censoring point using the strategy described in in Section \ref{sec:QCR_extrapolation}. Finally we compared the predicted coefficients with those obtained from fitting a standard quantile regression model to the uncensored VSE data. Exemplary results for one of the imputation cells are shown in Figure
\ref{fig:coeff-stability-med-qual-vse}.

The extrapolated quantile path often differs substantially from the path estimated using the full data sometimes leading to opposite signs. Such differences would lead to major changes in the imputed values. Since the quantile path of the coefficients are comparable for the BeH (see Figure \ref{fig:coeff-stability-med-qual-beh} in the appendix) and we observed similar effects for many of the different imputation cells (results can be obtained from the authors upon request), we abandoned this strategy and only focused on the remaining approaches for wage imputation.

\subsection{Evaluation of the different imputation strategies}\label{sec:eval_imp}
In this section, we compare three different imputation strategies: the classical Tobit imputation model as proposed for example in \citet{gartner2005imputation}, the doubly censored linear regression model that also introduces artificial left-censoring (at the 10th or 20th quantile), and the CQR approach that uses the estimated coefficients close to the censoring point. 

\begin{figure}
    \caption{Densities for different imputation strategies for selected imputation cells of the BeH.} 
    \label{fig:densities-various-imputations}
      \includegraphics[width=0.34\textwidth]{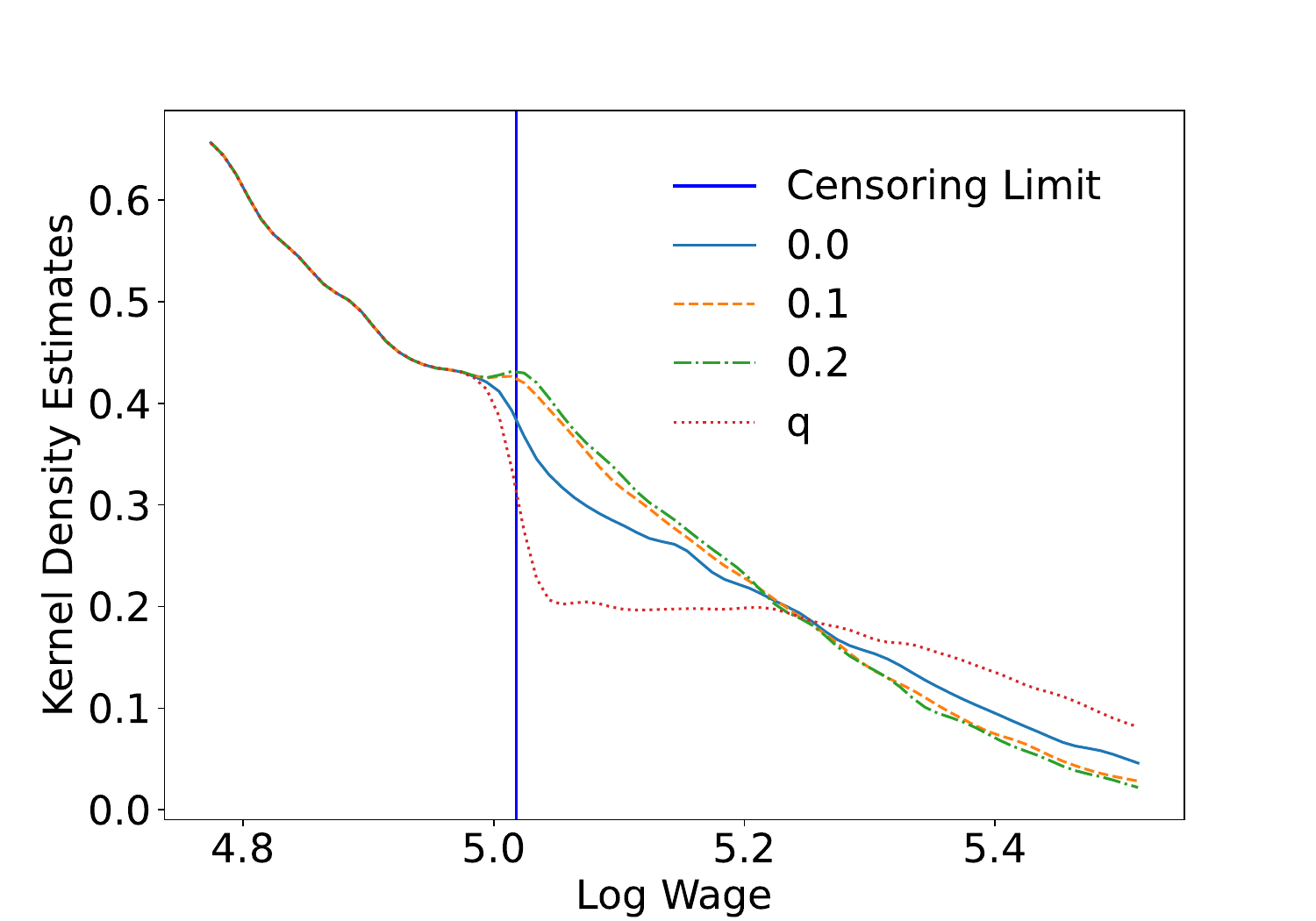}\hspace{-0.2cm}
      \includegraphics[width=0.34\textwidth]{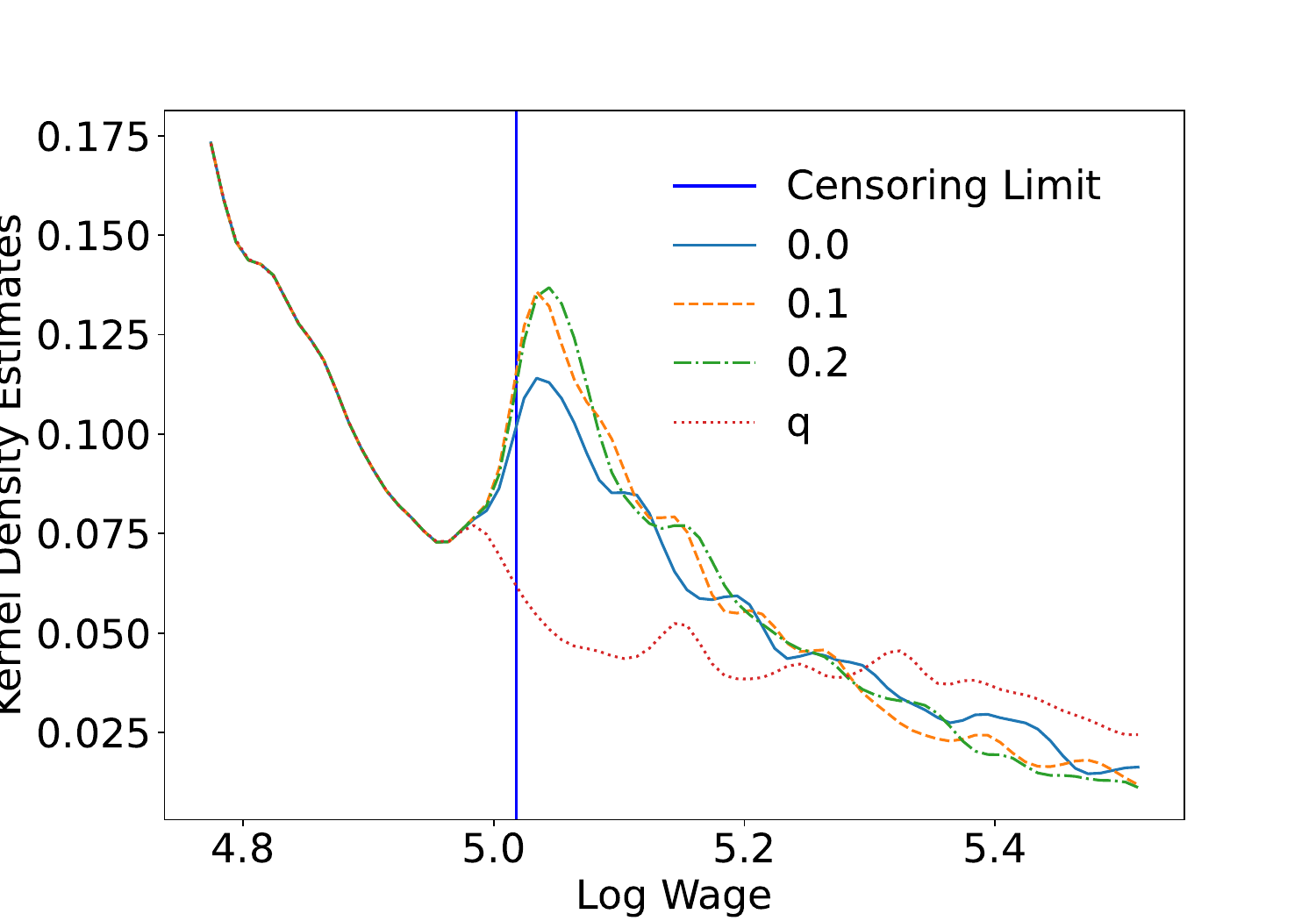}\hspace{-0.2cm}
      \includegraphics[width=0.34\textwidth]{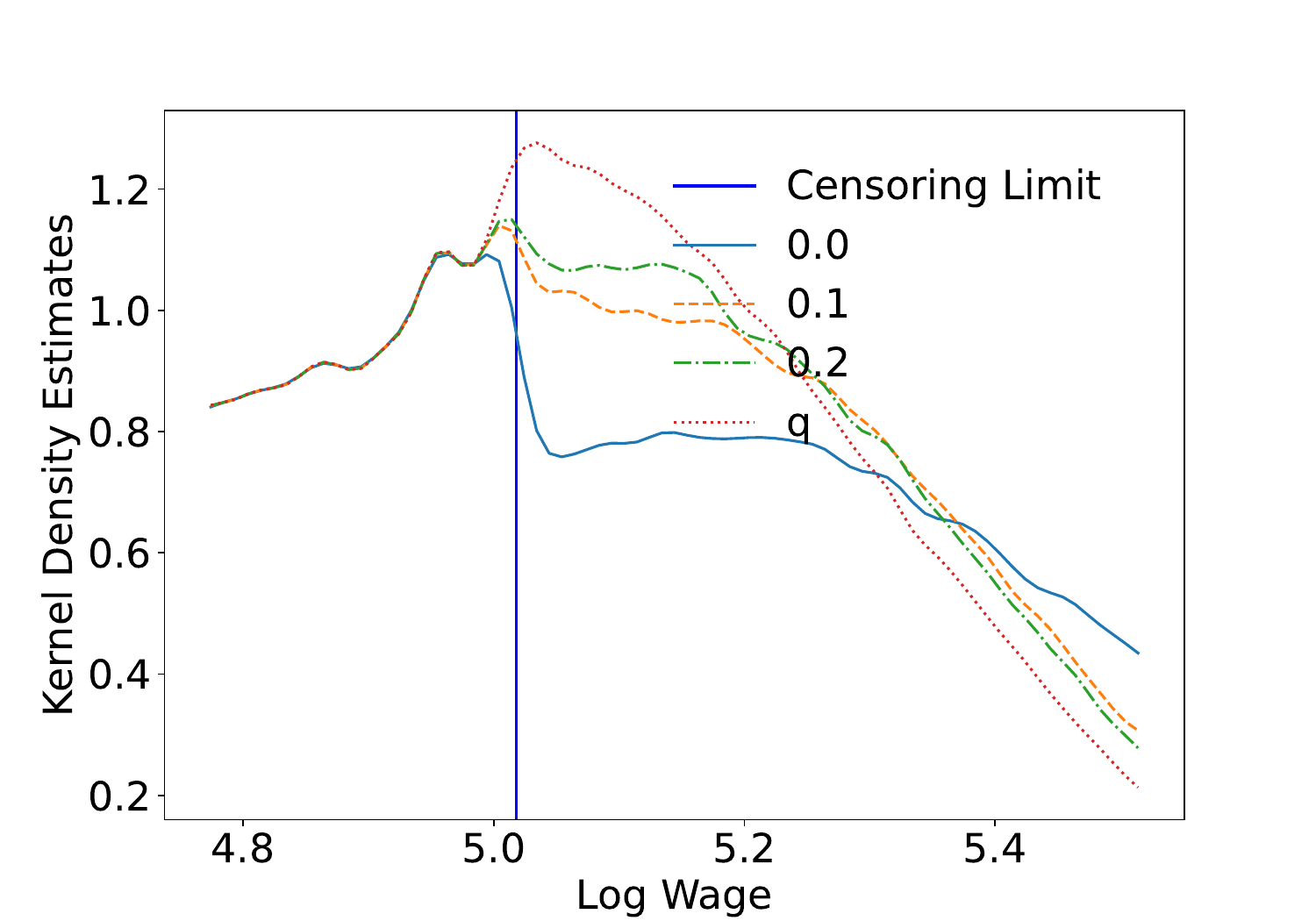}
    {\footnotesize Source: Own computations

  Data Source BeH, year=2010, subsamples: Medium-qualified men (completed apprenticeship),
    working in Western Germany, age group 45 to 64 years (left panel); Low-qualified men,
    working in Western Germany, age group 30 to 44 years (middle panel), High-qualified men,
    working in Western Germany, age group 30 to 45 years (right panel).}
\end{figure}

Results from the different imputation approaches for three of the 1,692 imputation cells used in analogy to the imputation strategy for the BeH are shown in Figure \ref{fig:densities-various-imputations}. 

We deliberately picked these three cells to illustrate that different imputation strategies seem to offer the most plausible densities for different cells (assuming that it is unreasonable to expect sudden jumps in the density around the censoring limit). In the left panel, the standard Tobit model seems to offer the best fit, the CQR approach seems preferable for the middle panel. Finally, doubly-censored quantile regression using the 20th quantile as an artificial censoring point seems to offer the best solution in the right panel. 

Given the inconsistent performance of the different imputation methods, a strategy is required to automatically identify the preferable imputation strategy, as manually reviewing the results for all 1,692 imputation cells is infeasible.
To address this problem, we developed a simple criterion that allows an automated ranking of the approaches.
The criterion measures the smoothness
of the kernel density around the censoring limit, computed as the sum
of the absolute values of the second finite derivatives of the kernel density (SAD), formally:
\[  \text{SAD}=\sum_{i \in G} \left|\frac{\Delta^2 \hat{f}(x_i)}{\Delta x_i^2}\right|, \]
where the  derivatives are evaluated for a grid $G = \{ g_{min}, g_{min} + 0.001, \ldots, g_{max}-0.001, g_{max}\}$ with $g_{min} = 0.99 \times C$, $g_{max} = 1.01 \times C$.\footnote{We considered another sensible (but in its original definition infeasible) criterion as the weighted (discrete) integral of the (absolute) deviation between the kernel density estimate  $f^j(x_i)$ of approach $j$ and the  density $f^u(x_i)$ of the uncensored wages (which are available for the VSE only). Formally
\[ \sum_{i \in G} \big| f^u(x_i) - f^j(x_i) \big| \times (x_i - x_{i-1}) \times
     f^u(x_i) .\]
This criterion becomes feasible by replacing the true density $ f^u(x_i)$
by the extrapolated $\hat{f}^u(x_i)$. We obtained  $\hat{f}^u(x_i)$ by approximating the  true density in the the interval $\{ g_{min}, g_{min} + 0.001, \ldots, C-0.001\}$ with $g_{min} = 0.9 \times C$ by a linear least squares regression and using the predicted values from the model $\hat{f}^u(x_i)$ in the comparison interval $G = \{ g_{min}, g_{min} + 0.001, \ldots, g_{max}\}$  Since both criteria yielded the same results in almost all situations, the second criterion was abandoned for sake of simplicity.}

We note that the evaluation metric depends on several parameters: the width of the window in which the distributional changes are evaluated, the coarseness of the grid, and the bandwidth for the kernel density estimator. We found in extensive evaluations with varying parameter combinations that the criterion is relatively robust to the exact parameter settings.

\begin{figure}
    \caption{Densites for all Imputation Approaches}
    \label{fig:densities-med-qual-vse}
\includegraphics[scale=0.2,trim=2cm 0cm 0cm 2cm,clip]{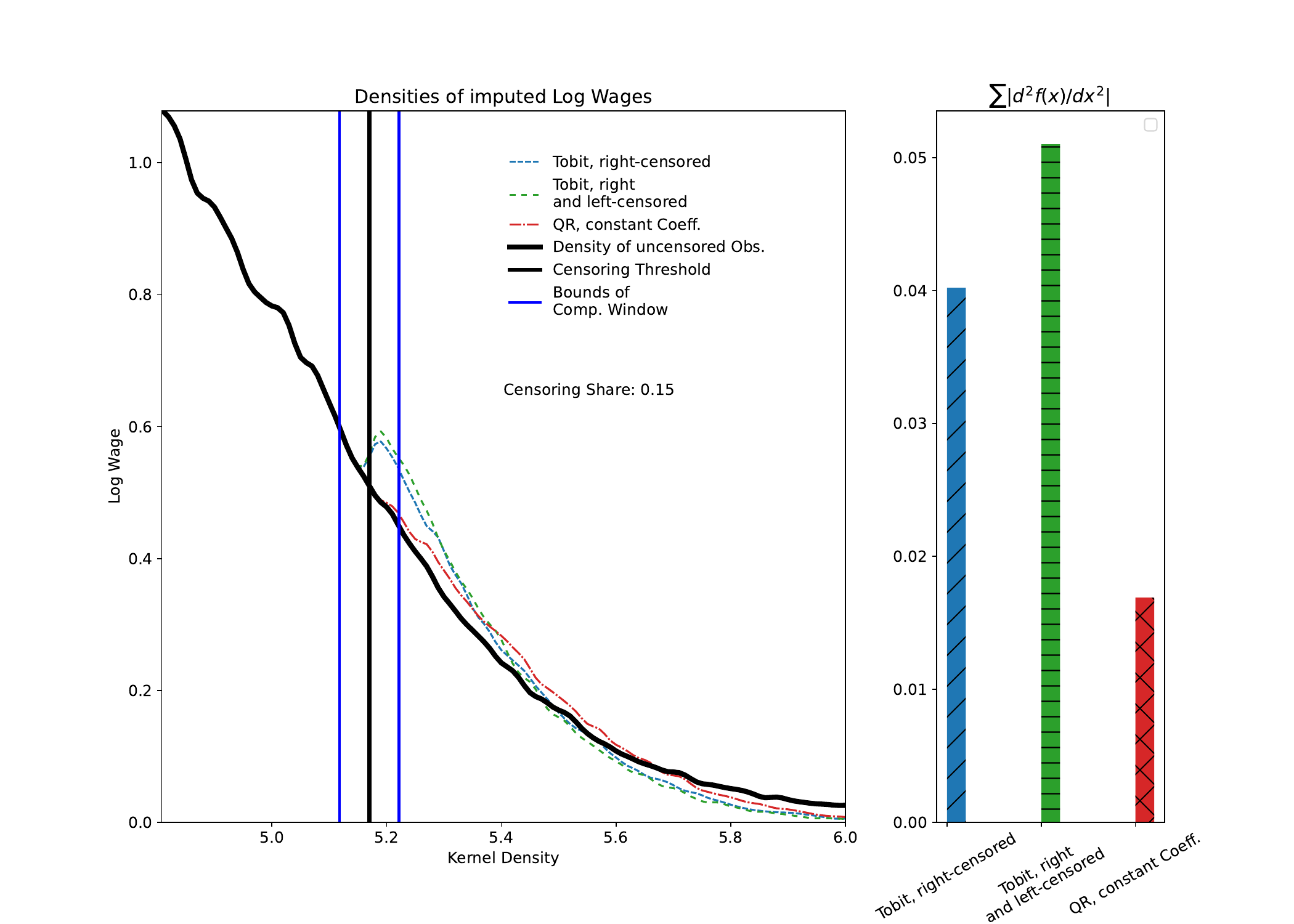}
  \includegraphics[scale=0.2,trim=2cm 0cm 0cm 2cm,clip]{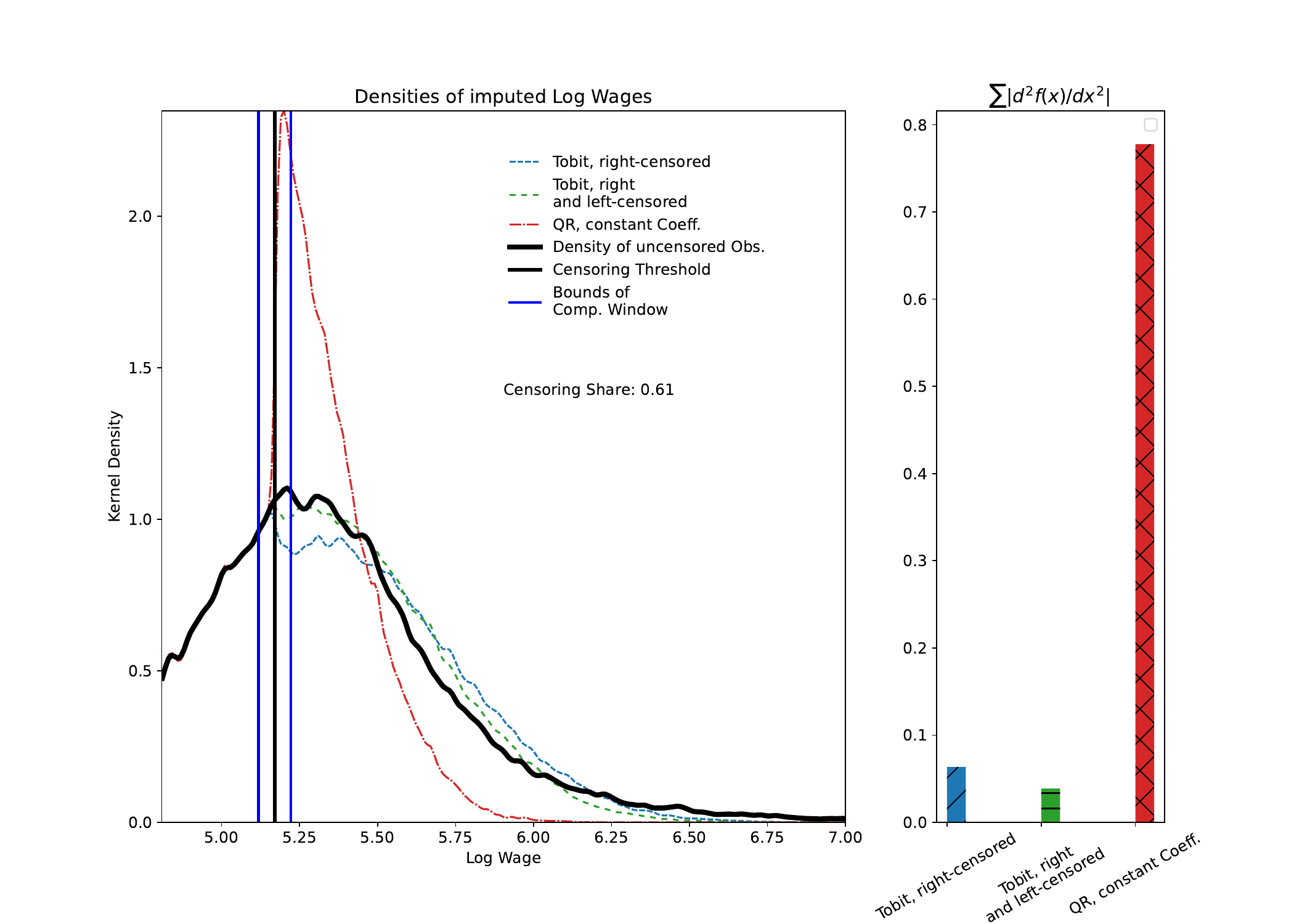}
    {\footnotesize Source: Own computations

  Data Source VSE 2010, subsample: Medium-qualified men (left panel)/highly-qualified men (right panel),
    working in Western Germany, age group 30 to 64 years.}
\end{figure}

To evaluate the usefulness of the criterion, we relied on the VSE again. Figure \ref{fig:densities-med-qual-vse} shows the results for two imputation cells. The black vertical line indicates the artificial censoring threshold that we introduced based on the threshold in the BeH. The blue vertical lines indicate the lower and the upper bound of the evaluation window. The solid black line indicates the density of the true uncensored data, while the dotted and dashed lines represent the densities based on the different imputation strategies. The bar charts next to the graphs show the results of the evaluation criterion for the different imputation approaches. The higher the bar the more pronounced are the distributional changes around the censoring limit, that is, the method with the smallest bar should be preferred. For both imputation cells, we find that the criterion picks the imputation strategy for which the density of the imputed values is closest to the true density distribution.

While Figure \ref{fig:densities-med-qual-vse} indicates that the marginal income distribution will be well preserved if the best fitting model according to the evaluation criterion is selected, it remains unclear whether the selected model will also provide the best results for other analysis tasks such as running regression models or computing unconditional statistics like means, variances and quantiles. We assess this by computing quality measures for various analysis tasks. Specifically, based on a regression model including a constant, a quadratic polynomial in age and the log establishment size, we evaluate the following statistics:

\begin{enumerate}
    \item The Mean Squared Error of prediction ($MSE_{pred}$), i.e. the mean of the squared
    difference between the predicted values from the least squares regression model based on the  true uncensored (log) wages and a model based on imputed wages.
    
    \item The Mean Absolute Error of prediction ($MAE_{pred}$), i.e. the mean of the absolute
    difference between the predicted values from the regression model based on the  true uncensored (log) wages and a model based on imputed wages.

    \item The mean of squared differences ($MSD_{coef}$) between the coefficients from a regression model using the imputed wages and the coefficients from a model
    using the uncensored wages.

    \item The mean of absolute differences ($MAD_{coef}$) between the coefficients from a regression model using the imputed wages and the coefficients from a model
    using the uncensored wages.
\end{enumerate}

To also assess the differences in the marginal distribution, we look at the difference between the true uncensored and imputed wages at 90th and 99th quantile of the respective wage distribution. Finally, to measure the difference over the entire marginal distribution, we also compute the Kullback-Leibler divergence (KL divergence) between the two distributions.  

\begin{table}[ht]

    \caption{Selection Criterion and Regression Quality Measures -- Subsample:  Men, 30-64 Years, Western Germany. Data Source: VSE 2010\hide{{\color{red}Ich habe in der Liste die Bezeichnungen der verschiedenen Kriterien leicht angepasst (z.B. $MAD_{coef}$) um ein bisschen sprechendere Namen zu haben. Kannst du das in der Tabelle auch anpassen, sobald die fertig ist?Du wolltest ja noch die KL-Divergenz ergänzen.}}}
    
    \footnotesize
    {\centering \begin{tabular}{lrrrrrr}
\toprule
Qualification & \multicolumn{3}{l}{Completed Apprenticeship} & \multicolumn{3}{l}{College} \\
Estimator &                  Tobit R & Tobit LR &     QR & Tobit R & Tobit LR &     QR \\
\midrule
$MSE_{pred}$           &                    0.162 &    0.162 &  0.162 &   0.176 &    0.176 &  0.191 \\
$MAE_{pred}$           &                    0.299 &    0.299 &  0.299 &   0.306 &    0.303 &  0.312 \\
$MSD_{coef}\times 100$ &                    0.001 &    0.001 &  0.001 &   0.032 &    0.021 &  0.038 \\
$MAD_{coef}\times 100$ &                    0.188 &    0.224 &  0.187 &   1.374 &    0.938 &  1.800 \\
$KL-div \times 100$    &                    0.147 &    0.179 &  0.096 &   0.470 &    0.489 &  1.976 \\
Dev. Q(90)            &                   -0.019 &   -0.025 & -0.002 &   0.013 &   -0.052 & -0.322 \\
Dev. Q(99)           &                   -0.264 &   -0.284 & -0.226 &  -0.386 &   -0.486 & -0.888 \\
SAD                   &                    0.044 &    0.065 &  0.024 &   0.056 &    0.044 &  0.783 \\
\bottomrule
\end{tabular}
 }

    Legend: $MSE_{pred}$ ($MAE_{pred}$): mean squared (absolute) deviations between predictions estimated
    with imputed wages and predictions obtained with true uncensored wages. $MSD_{coef}$, ($MAD_{coef}$): mean squared (absolute) deviations between coefficients estimated
    with imputed wages and coefficients obtained with true uncensored wages.
    $KL-div$: Kullback-Leibler divergence. Dev. $Q(q)$: Deviation between quantile $q$ of uncensored and imputed wages,
    SAD: Sum of absolute second finite derivatives.
    \label{tab:criterion-sex1-east0}
\end{table}

Table \ref{tab:criterion-sex1-east0} shows the results for medium and high-qualified men working in Western Germany. The tables for women and Eastern Germany which yield qualitatively similar results are included in Appendix \ref{sec:further_qual_results}.

The last row contains the selection criterion SAD. It favors the quantile regression for
the medium qualified (completed apprenticeship) and the doubly-censored Tobit for 
the high qualified (college graduates). Note that the ordering of all quality measures relating to regression models ($MSE_{pred}$, $MAE_{pred}$, $MSD_{coef}$ and $MAD_{coef}$) is identical to the ordering according to the SAD, implying that the SAD selects the imputation  model which yields best results if the imputed data are used for regression models. 

Looking at the deviations in the marginal distribution, we find that the results are less clear cut. For the medium qualified the favoured quantile regression yields the smallest deviation for all statistics. But this does not apply for the high-qualified where the doubly-censored Tobit model favored by the SAD criterion shows greater quantile deviations and KL divergence than the right-censored Tobit model for the two quantiles. Still, we note that the differences tend to be small and the SAD criterion correctly identifies that the quantile regression imputation is clearly inferior to the other two approaches for this subgroup. 

Similar results are obtained for the other imputation cells presented in the appendix. The SAD criterion tends to correctly identify the best imputation approach if the goal is to compute regression models on the imputed data. For the quantiles, the results are somewhat mixed. Obviously, the quantile comparisons are only snapshots at specific points in the distribution that could easily be affected by small kinks in the distribution. Using the KL divergence to assess how well the entire marginal distribution is preserved, we find that the SAD criterion selects the best performing imputation strategy in 6 out of 8 subsamples. The KL divergence would imply a different optimal imputation strategy for the male and female college graduates in Western Germany (Tables \ref{tab:criterion-sex1-east0}, \ref{tab:criterion-sex2-east0}).\hide{{\color{red} zumindest hoffe ich das...}. Tja, das stimmt leider nur in 6 von 8 Fällen. Damit können wir aber hoffentlich leben.} 

\subsection{Evaluations for the BeH}\label{sec:BEH_evaluations}
The evaluations presented so far relied on the VSE. The VSE has the important advantage that ground truth data are available since the income information is (almost) uncensored. However, the VSE is only cross-sectional and thus the LOOMs cannot be included in the imputation model and thus the results in Section \ref{sec:evaluation} are only generalizable to the longitudinal context if we assume that the LOOMs do not have any effect on the findings. 

To circumvent this strong assumption, we directly compare the marginal income distribution of the harmonized BeH (imputed using the LOOMs) with the marginal distribution of the VSE. We only focus on those subsamples from 2010 for which the densities below the censoring limit as sufficiently similar: Males, aged 30--64 living in Western Germany with medium or high qualification (see Figure \ref{fig:comp-densities-beh-vse} above). \footnote{The marginal distributions of the VSE and the BeH differ too much for all other subsamples in to allow a reliable comparison. Figures for all subsamples are available in an online appendix.}For these two subgroups, we compare our optimized imputation strategy with a classical approach based on the Tobit model.
\begin{figure}
    \caption{Comparison of the optimum (mixed) imputation strategy with a standard (right-censored) Tobit imputation model.}
    \label{fig:comp-opt-tobit}

\includegraphics[scale=0.37]{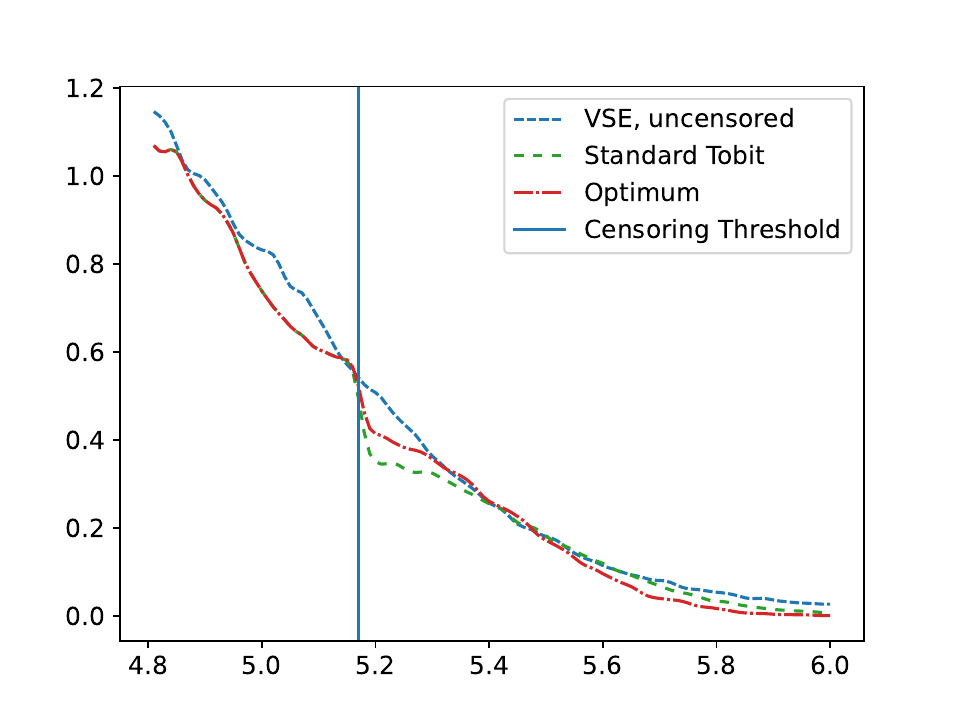}
\includegraphics[scale=0.37]{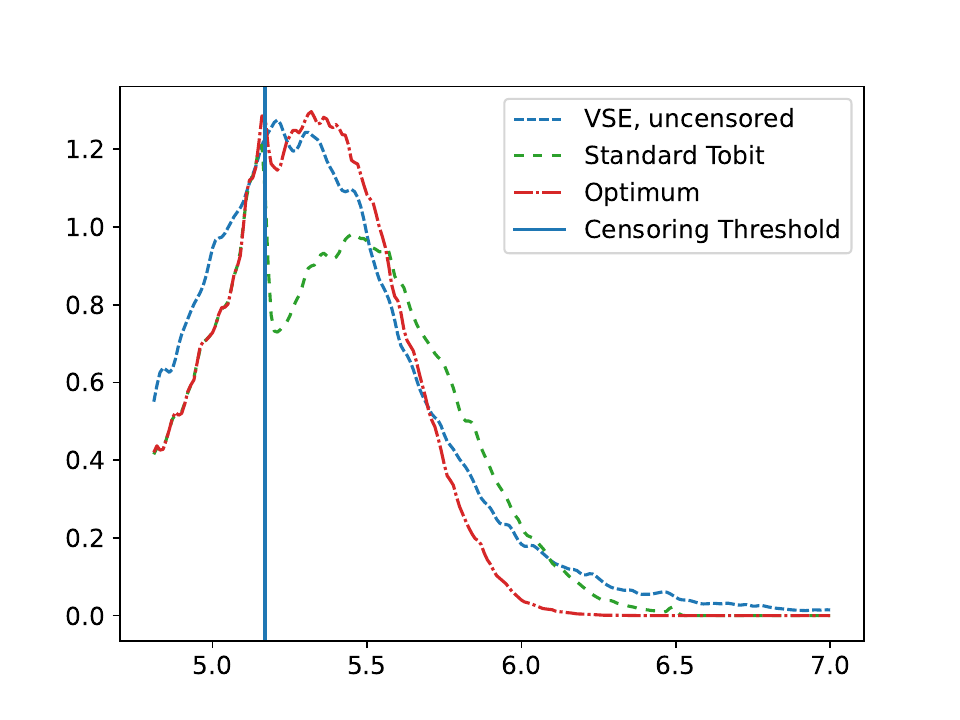}
  
    {\footnotesize Source: Own computations

  Data Sources VSE and BeH 2010, subsamples: Medium-qualified men (left panel)/highly-qualified men (right panel), working in Western Germany, age group 30 to 64 years.}
\end{figure}

Figure \ref{fig:comp-opt-tobit} shows the marginal distribution of the two data sources. The respective KL divergences are listed in Table \ref{tab:comp-opt-tobit}.  Two aspects are noteworthy: (1) The two distributions match closely up to the censoring threshold. Based on the assumption that this is also true for the unobserved part of the income distribution of the BeH above the censoring limit, we can compare the observed income distribution of the VSE with the distribution of the imputed income above the censoring threshold. (2) The distribution of the imputed income, which resulted from always picking the optimal imputation procedure according to the SAD criterion in each imputation cell, follows the distribution of the VSE more closely than the distribution based on a standard right-censored Tobit, indicating a higher imputation quality. \hide{{\color{red} Wäre natürlich schön, wenn wir das mit einer reinen Tobitimputation vergleichen könnten, um hoffentlich zu zeigen, dass unsere Imputation tatsächlich zu einer deutlichen Verbesserung führt}. Hab ich jetzt eingefügt.} This is confirmed by the KL divergences shown in Table \ref{tab:comp-opt-tobit}: The divergences are always smaller for the optimal imputation method. For the highly-qualified the divergence is only half of the divergence when using the Tobit imputation model.

\begin{table}
    \caption{KL divergences for the two imputation approaches.}
    \label{tab:comp-opt-tobit}
\begin{center}   \begin{tabular}{lrrr}
\toprule
Qualification &   KLD Optimum &  KLD Tobit &  Relative KLD \\
              &               &            & (Optimum / Tobit)  \\
\midrule
 medium-qualified &        0.016 &      0.018 &         0.870 \\
 highly-qualified &        0.074 &      0.156 &         0.477 \\
\bottomrule
\end{tabular}
 \end{center}
\end{table}

\hide{Figure XX compares the distributions for the different imputation strategies in some of the imputation cells. Each panel also reports the SAD and the Kullback-Leibler divergence. We see that the SAD seems to be a reliable criterion for picking the optimal imputation procedure. In X of the Y panels, the criterion picks the imputation procedure with the smallest Kullback-Leibler divergence. In fact, the optimal imputation procedure according to the KL-divergence is picked in ZZ\% of the 36 (??) imputation cells used in 2010.}  

\section{Conclusion}
The goal of the project that motivated the research presented in this paper was to obtain imputed wages for the German employment register data (BeH) that can be used for various downstream analyses tasks. We achieved this by including (proxies for) fixed effects at several levels and estimating the models separately for narrow cells based on year, age, gender and education groups. Inspection of the densities of the imputed wages based on the classical Tobit model revealed substantial kinks and bumps at the censoring threshold. \hide{Since the models are highly parameterized, omitted regressors are unlikely to be the cause of these deficiencies.} We identified the variability of the regression coefficients as a function of the quantiles of the income distribution as the likely cause of these deficiencies and tackled the problem by either introducing artificial left-censoring of the wages or using quantile regressions in order to obtain the relevant `local' regression coefficients. As none of the approaches strictly dominated the others, we proposed a selection criterion based on the smoothness of the imputed income distribution around the censoring limit. Extensive simulation studies using the VSE as test bed illustrate the suitability of this selection criterion.

The proposed modeling approach appears useful in two respects. First, it is applicable for a wide class of right-censored variables. Second, it offers a specification test for Tobit models which should be conducted even if generating imputations seems unnecessary. Since Tobit models are based on the assumption of constant regression coefficients they yield biased results if this assumption is violated. Computing imputed values and inspecting their density at the censoring threshold may therefore uncover dependency of the regression coefficients on the quantiles of the dependent variable. \hide{{\color{red} Ginge das nicht auch einfach mit einer zensierten Quantilsregression? Klar. Nicht unbedingt. Manchmal sieht es ja so aus, als würden die QR-Koeffizienten sehr gut passen und das Tobit daneben liegen und trotzdem sieht die Dichte der Tobit-Imputation dann plausibler aus.}}

We note that it would be straightforward to generate multiple imputations based on the procedures outlined in this paper by simply drawing multiple values for each missing value. We did not pursue this for two reasons: Given the large size of the data containing more than two billion records, generating multiple copies of the imputed data requires a significant amount of extra storage. But more importantly, including the LOOMs in the imputation model leads to regression models with very high predictive power. As a consequence the model uncertainty that the multiple imputation approach tries to take into account, will be very small. Since the BeH covers the entire population, that is, there is no sampling uncertainty, any effects that will not be statistically significant because of the modelig uncertainty, are unlikely to be practically significant to begin with.
\hide{}
\bibliography{bibliography.bib}
\bibliographystyle{chicago}
\clearpage 
\begin{appendix}
\section{Formal definitions of the leave-one-out-means}
\label{sec:appendix-formulas}
Here we provide precise formal definitions of the person-specific ($\tilde{\mu}_{i, -s}$) and 
establishment-specific ($\tilde{\eta}_{-i,et}$) LOOMs.\footnote{The definition of the occupation-specific LOOMs is omitted since it can be obtained by replacing the establishment index by the occupation index in the establishment-specific LOOMs.}
\begin{align} \label{eq:loo-def}
  \tilde{\mu}_{i, -s} & = \ln\left(\frac{1}{n_{i,-s}} \sum_{s' \neq s} W_{s'i} \, d_{s'i} \right)\\
  \tilde{\eta}_{-i,et} & = \ln\left(\frac{1}{n_{-i,et}} \sum_{t' \in \theta(t)} \sum_{s\neq s'} \sum_{j \neq i} W_{s'jet'} \, d_{s'jet'} \right),
\end{align}
where $W_{siet}$ denotes the daily wage, $d_{si(et)}$ denotes the duration of spells and $\theta(t)$ denotes the set of years which are averaged. Specifically,
\[\theta(t) = \left\{\begin{array}{cc}
                       \{t, t+1\}, & \text{if establishment  $e$ is founded in $t$} \\
                         \{t-1, t\} & \text{if establishment $e$ is closed in $t$} \\
                       \{t-1, t, t+1 \} & \text{otherwise}
                     \end{array} \right. \]

Finally,
\begin{align*}
n_{i,-s} & =  \sum_{s' \neq s} \, d_{s'i}, \\
 n_{-i,et}& = \sum_{t' \in \theta(t)} \sum_{s'\neq s} \sum_{j \neq i} d_{s'jet'}
\end{align*}
denote the respective leave-one-out sums of spell durations.
\hide{\section{Comparison of the BeH with the VSE{\color{red} Brauchen wir das noch? Mir ist nicht ganz klar, welche Zusatzinfo das bietet.}
Antwort: hab vergessen, den Abschnitt aus dem Appendix zu löschen nachdem
ich ihn in den Hauptteil verschoben habe}
The VSE is used as a test bed in order to evaluate the imputation models proposed in this paper. Clearly it is suitable for this purpose only if the samples and the available variables are sufficiently similar. Here we describe the main characteristics of the VSE and demonstrate the
similarity of both data sets by comparing the frequencies of their age structures and the densities of wages for the subsamples which are used in the figures and tables in the main section 2 (Men, aged 30-64 years, working in Western Germany).}

\section{Results for the censored quantile regression based on the BeH}

The following Figure \ref{fig:coeff-stability-med-qual-beh} reproduces Figure \ref{fig:coeff-stability-med-qual-vse} based on the BeH. The coefficient profiles are highly similar to the respective VSE profiles. 
\begin{figure}
    \caption{Profiles of the Coefficients from Censored Quantile Regressions, Obtained from the BeH}
    \label{fig:coeff-stability-med-qual-beh}
    
    \includegraphics[width=0.99\textwidth, trim=2cm 3cm 2cm 0cm,clip]{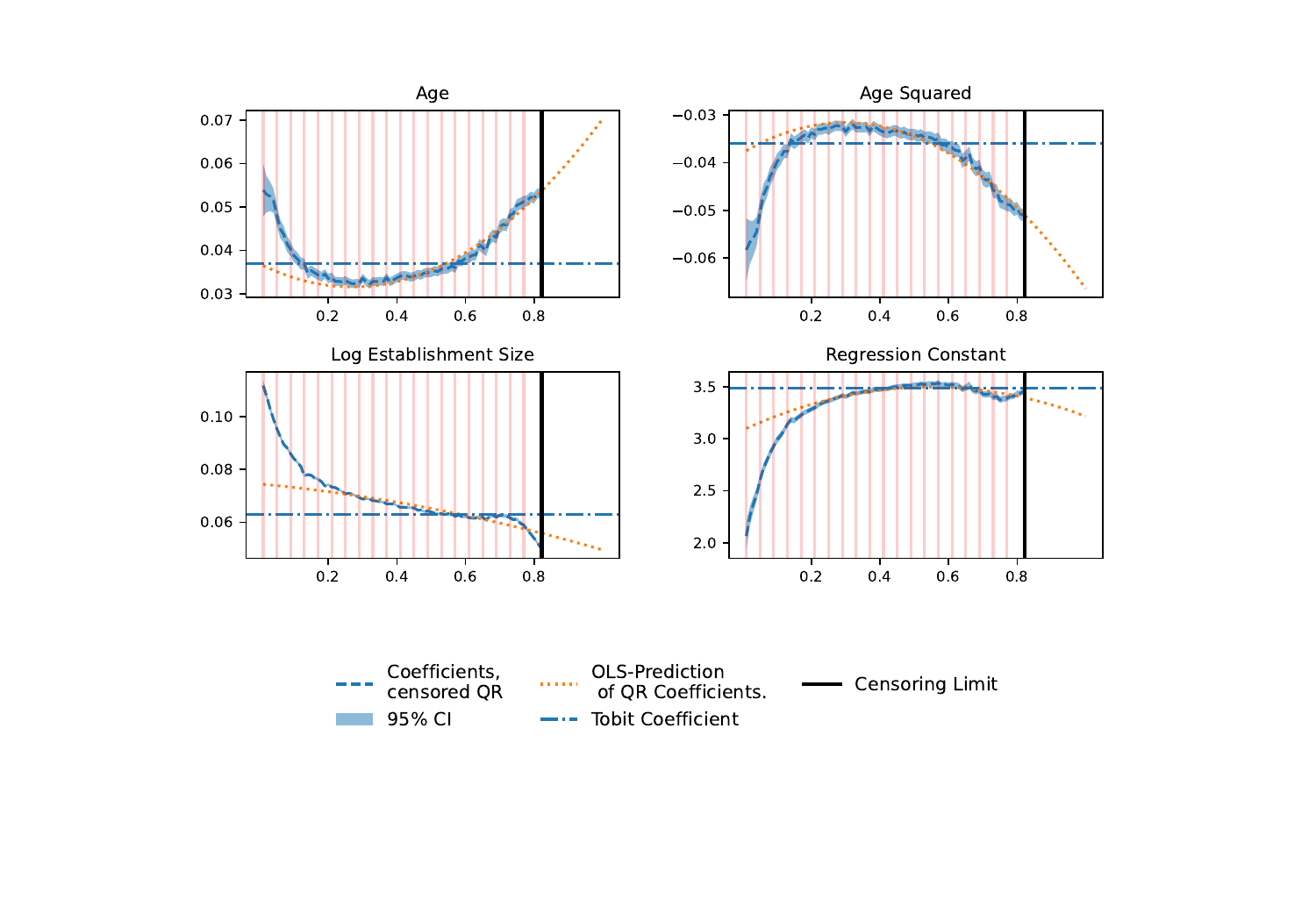}
    {\footnotesize Source: Own computations

  Data Source BeH, subsample: Medium-qualified men (completed apprenticeship),
    working in western Germany, age group 30 to 64 years.}
\end{figure}

\clearpage

\section{Further Tables for Assessing the Imputation Selection Criterion (SAD)}\label{sec:further_qual_results}

\begin{table}[ht]

    \caption{Selection Criterion and Regression Quality Measures -- Subsample:  Men, 30-64 Years, Eastern Germany. Data Source: VSE 2010}
    
    \footnotesize
    {\centering \begin{tabular}{lrrrrrr}
\toprule
Qualification & \multicolumn{3}{l}{Completed Apprenticeship} & \multicolumn{3}{l}{College} \\
Estimator &                  Tobit R & Tobit LR &     QR & Tobit R & Tobit LR &     QR \\
\midrule
$MSE_{pred}$           &                    0.162 &    0.162 &  0.162 &   0.206 &    0.207 &  0.213 \\
$MAE_{pred}$           &                    0.306 &    0.306 &  0.306 &   0.330 &    0.330 &  0.333 \\
$MSD_{coef}\times 100$ &                    0.000 &    0.000 &  0.000 &   0.005 &    0.004 &  0.030 \\
$MAD_{coef}\times 100$ &                    0.055 &    0.065 &  0.072 &   0.645 &    0.655 &  1.635 \\
$KL-div \times 100$    &                    0.139 &    0.135 &  0.119 &   0.793 &    0.704 &  1.332 \\
Dev. Q(90)            &                    0.000 &    0.000 &  0.000 &  -0.025 &   -0.061 & -0.234 \\
Dev. Q(99)           &                   -0.206 &   -0.191 & -0.109 &  -0.380 &   -0.447 & -0.765 \\
SAD                   &                    0.037 &    0.024 &  0.019 &   0.074 &    0.050 &  0.361 \\
\bottomrule
\end{tabular}
 }

    Legend: MSD (MAD) Coeff: Mean Squared (Absolute) Deviations of coefficients,
    Dev. Q $q$ Deviation between quantile $q$ of uncensored and imputed wages,
    SAD2: Sum of Absolute second finite Derivatives.
    \label{tab:criterion-sex1-east1}
\end{table}

\begin{table}[ht]

    \caption{Selection Criterion and Regression Quality Measures -- Subsample:  Women, 30-64 Years, Western Germany. Data Source: VSE 2010}
    
    \footnotesize
    {\centering \begin{tabular}{lrrrrrr}
\toprule
Qualification & \multicolumn{3}{l}{Completed Apprenticeship} & \multicolumn{3}{l}{College} \\
Estimator &                  Tobit R & Tobit LR &     QR & Tobit R & Tobit LR &     QR \\
\midrule
$MSE_{pred}$           &                    0.146 &    0.146 &  0.145 &   0.164 &    0.164 &  0.164 \\
$MAE_{pred}$           &                    0.288 &    0.288 &  0.288 &   0.299 &    0.299 &  0.299 \\
$MSD_{coef}\times 100$ &                    0.000 &    0.000 &  0.000 &   0.005 &    0.006 &  0.009 \\
$MAD_{coef}\times 100$ &                    0.087 &    0.089 &  0.088 &   0.674 &    0.804 &  0.931 \\
$KL-div \times 100$    &                    0.106 &    0.128 &  0.135 &   0.539 &    0.523 &  0.472 \\
Dev. Q(90)            &                    0.000 &    0.000 &  0.000 &  -0.007 &   -0.040 & -0.055 \\
Dev. Q(99)           &                   -0.097 &   -0.127 & -0.001 &  -0.205 &   -0.287 & -0.345 \\
SAD                   &                    0.027 &    0.031 &  0.036 &   0.036 &    0.039 &  0.063 \\
\bottomrule
\end{tabular}
 }

    Legend: MSD (MAD) Coeff: Mean Squared (Absolute) Deviations of coefficients,
    Dev. Q $q$ Deviation between quantile $q$ of uncensored and imputed wages,
    SAD2: Sum of Absolute second finite Derivatives.
    \label{tab:criterion-sex2-east0}
\end{table}

\begin{table}[ht]

    \caption{Selection Criterion and Regression Quality Measures -- Subsample:  Women, 30-64 Years, Eastern Germany. Data Source: VSE 2010}
    
    \footnotesize
    {\centering \begin{tabular}{lrrrrrr}
\toprule
Qualification & \multicolumn{3}{l}{Completed Apprenticeship} & \multicolumn{3}{l}{College} \\
Estimator &                  Tobit R & Tobit LR &     QR & Tobit R & Tobit LR &     QR \\
\midrule
$MSE_{pred}$           &                    0.172 &    0.172 &  0.172 &   0.166 &    0.166 &  0.167 \\
$MAE_{pred}$           &                    0.332 &    0.332 &  0.332 &   0.312 &    0.312 &  0.312 \\
$MSD_{coef}\times 100$ &                    0.000 &    0.000 &  0.000 &   0.000 &    0.000 &  0.001 \\
$MAD_{coef}\times 100$ &                    0.074 &    0.065 &  0.043 &   0.136 &    0.145 &  0.236 \\
$KL-div \times 100$    &                    0.099 &    0.098 &  0.135 &   0.754 &    0.776 &  0.723 \\
Dev. Q(90)            &                    0.000 &    0.000 &  0.000 &  -0.009 &   -0.006 & -0.037 \\
Dev. Q(99)           &                   -0.039 &   -0.040 &  0.058 &  -0.149 &   -0.151 & -0.279 \\
SAD                   &                    0.026 &    0.011 &  0.031 &   0.122 &    0.098 &  0.094 \\
\bottomrule
\end{tabular}
 }

    Legend: MSD (MAD) Coeff: Mean Squared (Absolute) Deviations of coefficients,
    Dev. Q $q$ Deviation between quantile $q$ of uncensored and imputed wages,
    SAD2: Sum of Absolute second finite Derivatives.
    \label{tab:criterion-sex2-east1}
\end{table}
\end{appendix}


\end{document}